\documentclass[prd,twocolumn,nofootinbib,superscriptaddress]{revtex4-2}
\usepackage{bm}
\usepackage{amsmath}
\usepackage{amssymb}
\usepackage{graphicx}
\usepackage{subfigure}
\usepackage{hyperref}
\usepackage{color}
\usepackage{comment}
\usepackage{ulem}
\usepackage{CJK}
\hypersetup{
	colorlinks=true,
	linkcolor=red,
	citecolor=blue,
}

\allowdisplaybreaks[2]

\usepackage{graphicx}
\usepackage{dcolumn}
\usepackage{bm}
\begin{document}

\preprint{APS/123-QED}

\title{Eccentricity Effects on Modeling Dynamic Quantities and Their Correlations in Binary Black Hole Mergers}

\author{Hao Wang}
\email{husthaowang@hust.edu.cn}
\affiliation{Department of Astronomy, School of Physics, Huazhong University of Science and Technology, Wuhan 430074, China}

\author{Yuan-Chuan Zou}
\email{zouyc@hust.edu.cn}
\affiliation{Department of Astronomy, School of Physics, Huazhong University of Science and Technology, Wuhan 430074, China}

\author{Qing-Wen Wu}
\email{qwwu@hust.edu.cn}
\affiliation{Department of Astronomy, School of Physics, Huazhong University of Science and Technology, Wuhan 430074, China}

\date{\today}

\begin{abstract}
In this study, we begin by revisiting the oscillatory behavior of radiative quantities—energy, angular momentum, and linear momentum—linked with initial eccentricities in binary black hole (BBH) mergers. By varying the mean anomaly $l_0$ across the parameter range $[0,2\pi]$ from a post-Newtonian perspective, we establish an envelope that encapsulates the oscillations of these radiative quantities. Our analysis reveals that while the oscillations are influenced by the specific initial condition $l_0$, the effect of eccentricity contributes to the formation of this envelope.
Subsequently, we model dynamical quantities such as peak luminosity $L_{\text{peak}}$, remnant mass $M_{\text{rem}}$, spin $\alpha_{\text{rem}}$, and recoil velocity $V_{\text{rem}}$ in circular orbits. Through polynomial modeling, we explore their relationships with mass ratios and correlations. Our results demonstrate the effectiveness of these polynomials in capturing the intricate relationships and correlations among these quantities in circular orbits.
Furthermore, we synthesize and analyze dynamical quantities for both circular and eccentric orbits, revealing continuous variations within specific ranges corresponding to distinct mass ratios. These variations are influenced by continuous changes in initial eccentricity and the associated envelope, which can be extrapolated to encompass other mass ratios. By interpolating the maximum and minimum values of these dynamical quantities, we unveil considerably broad domains relative to circular orbits in both orbital and non-orbital BBH mergers. These domains provide robust constraints on the relationships between dynamical quantities, mass ratios, and their correlations.
Finally, we discuss the extension of this eccentricity effect to spin alignment and spin precession configurations of BBHs.

\end{abstract}
\maketitle

\section{Introduction}
The epoch of regular gravitational wave detection began in 2015, signifying a momentous leap forward in astrophysical observations \cite{LIGOScientific:2016aoc}. Currently, the ground-based gravitational wave detectors, including LIGO \cite{LIGOScientific:2014qfs}, Virgo \cite{VIRGO:2014yos}, and KAGRA \cite{KAGRA:2018plz}, have collectively detected nearly one hundred binary compact object mergers \cite{LIGOcollaboration}.

The primary focus of contemporary gravitational wave detections pertains to circular orbit BBH mergers, primarily attributed to the circularization effect induced by gravitational radiation \cite{Peters:1963ux,Peters:1964zz}. Nonetheless, within star-dense environments like globular clusters \cite{Miller:2002pg,Gultekin:2005fd,OLeary:2005vqo,Rodriguez:2015oxa,Samsing:2017xmd,Rodriguez:2017pec,Rodriguez:2018pss,Park:2017zgj} and galactic nuclei \cite{Gondan:2020svr,Antonini:2012ad,Kocsis:2011jy,Hoang:2017fvh,Gondan:2017wzd,Samsing:2020tda,Tagawa:2020jnc}, dynamical interactions involving two-body \cite{East:2012xq}, three-body \cite{Naoz:2012bx,VanLandingham:2016ccd,Silsbee:2016djf,Blaes:2002cs,Antognini:2013lpa,Stephan:2016kwj,Katz:2011hn,Seto:2013wwa}, four-body \cite{Zevin:2018kzq} systems, and beyond, can instigate the formation of BBH systems with eccentric orbits. The origin of circular orbit BBH mergers, whether arising from isolated binary evolution or dynamic interactions, remains ambiguous. However, the identification of an eccentric orbit BBH merger undeniably indicates the formation of a dynamically evolved system.
The event GW190521 \cite{LIGOScientific:2020iuh} stands as a notable example, being identified as a probable outcome of an eccentric orbit BBH merger \cite{Gayathri:2020coq,Romero-Shaw:2020thy}, indicative of dynamic formation processes at play.

Numerous techniques exist for resolving the dynamics of BBH mergers, including Effective One Body (EOB) \cite{Buonanno:1998gg,Damour:2001tu}, post-Newtonian (PN) \cite{Blanchet:2013haa}, black hole perturbation theory \cite{Teukolsky:1973ha}, and numerical relativity (NR) \cite{Pretorius:2005gq}. Among these, NR stands out as the most precise method capable of fully elucidating the strong field dynamics of BBH systems. From NR simulations, we derive essential information like gravitational waveforms and dynamic quantities such as peak luminosity, mass, spin, recoil velocity, etc., which serve as direct inputs for constructing gravitational wave templates and modeling the properties of remnant black hole in astrophysical contexts. 

Modeling dynamical quantities involves establishing connections between the initial parameters and final parameters of BBH systems through various methodologies. Over the last decade, NR has made significant strides, progressing from initial equal-mass nonspinning BBH configurations to encompass spin-aligned, spin-precessing, and unequal-mass BBH systems, and evolving from circular orbit BBH mergers to include eccentric orbits \cite{Duez:2018jaf,Campanelli:2005dd,Baker:2005vv}. Several NR collaborations, such as Simulating eXtreme Spacetimes (SXS) \cite{Mroue:2013xna,Boyle:2019kee}, Rochester Institute of Technology (RIT) \cite{Healy:2017psd,Healy:2019jyf,Healy:2020vre,Healy:2022wdn}, bi-functional adaptive mesh (BAM) \cite{Hamilton:2023qkv,Bruegmann:2006ulg,Husa:2007hp}, and MAYA \cite{Jani:2016wkt,Ferguson:2023vta}, have conducted extensive simulations across these diverse scenarios. These findings establish favorable conditions for modeling dynamic quantities.

During this period, numerous studies have endeavored to model the relationships between the initial parameters of NR simulated BBH systems and crucial dynamical quantities like peak luminosity $L_{\text{peak}}$, mass $M_{\text{rem}}$, spin $\alpha_{\text{rem}}$, and recoil velocity $V_{\text{rem}}$, employing a range of different approaches. However, the bulk of these modeling efforts have predominantly focused on circular orbit BBH mergers. For instance, some Refs. \cite{Hofmann:2016yih,Lousto:2009mf,Buonanno:2007sv,Lousto:2009ka,Rezzolla:2007rz} have correlated the remnant mass and spin with the energy and angular momentum of the innermost stable circular orbit and mass ratio, while others have used PN \cite{Blanchet:2005rj}, EOB \cite{Damour:2006tr}, and closed limit approximation \cite{Sopuerta:2006wj} methods to model the remnant recoil velocity. Additionally, certain studies have employed polynomial fitting of initial spin and mass ratio to model the remnant mass, spin, recoil velocity, and peak luminosity \cite{Taylor:2020bmj,Healy:2016lce,Healy:2014yta,Lousto:2013wta,Zlochower:2015wga,Jimenez-Forteza:2016oae,Rezzolla:2007xa,Ferguson:2019slp,Koppitz:2007ev,Campanelli:2007cga,Gonzalez:2006md,Hemberger:2013hsa,Carullo:2023kvj}, while others \cite{Varma:2018aht} have utilized non-analytical methods like Gaussian process regression to model the remnant mass, spin, and recoil velocity.

There has been a notable absence of modeling efforts focused on dynamical quantities in BBH mergers with eccentric orbits. This scarcity primarily stems from the lack of extensive NR simulations for BBH mergers in eccentric orbits. However, in recent years, collaborations like SXS, MAYA, and RIT have expanded their BBH simulations to include eccentric orbits.

In terms of modeling dynamical quantities in eccentric BBH mergers, only a single Ref. \cite{Islam:2021mha} has attempted to utilize Gaussian process regression to model the mass and spin of remnants based on twenty sets of eccentric orbit waveforms from SXS. However, this modeling effort is hindered by a limited sample size, and due to the relatively low eccentricities in these NR simulations, the properties of the remnants closely resemble those from circular orbits, lacking universal significance.

Notably, the fourth data release \cite{Healy:2022wdn} from RIT offers a wealth of data with sufficiently dense initial eccentricities and a broad coverage of parameter spaces in NR simulations. This dataset \cite{RITBBH} presents a valuable opportunity for in-depth exploration of the characteristics of dynamical quantities in BBH mergers with eccentric orbits.

In our previous work \cite{Wang:2023vka}, we established a comprehensive relationship between the dynamical quantities and initial eccentricity for eccentric orbit BBH mergers, utilizing all available eccentric orbit BBH merger data from RIT. Our findings revealed a gradual oscillation in the dynamical quantity as the initial eccentricity increases, eventually reaching extrema at specific initial eccentricity values. Subsequently, as BBH systems transition from orbital mergers to non-orbital mergers, the dynamical quantity stabilizes or tends towards zero.

In another study \cite{Wang:2024jro,Wang:2024afj}, we compared eccentric PN waveforms with eccentric NR waveforms in the inspiral phase, identifying that the oscillation in dynamical quantities is attributed to the non-orbital averaging effect, specifically the influence of the mean anomaly $l$, which is unique in eccentric waveforms.

In this study, we delve into modeling relationships of dynamical quantities such as peak luminosity $L_{\text{peak}}$, remnant mass $M_{\text{rem}}$, spin $\alpha_{\text{rem}}$, and recoil velocity $V_{\text{rem}}$ with mass ratio and their correlations in eccentric BBH mergers relative to the dynamical quantities in circular orbits from a more comprehensive and profound perspective.

This article is structured as follows. In Sec. \ref{sec:II}, we introduce the NR data of eccentric BBH mergers utilized in Sec. \ref{sec:II:A} and the eccentric PN method discussed in Sec. \ref{sec:II:B}. Sec. \ref{sec:II:C} explores how initial mean anomalies form the radiative quantities' envelopes based on the initial eccentricity. Polynomial modeling of dynamical quantities for circular BBH mergers is presented in Sec. \ref{sec:II:D}, while in Sec. \ref{sec:II:E}, we combine and analyze the dynamical quantities of both circular and eccentric BBH mergers.
In Sec. \ref{sec:III}, we delineate the domains shaped by the dynamical quantities of eccentric orbital BBH mergers concerning circular BBH mergers in Sec. \ref{sec:III:A}, and in Sec. \ref{sec:III:B}, we depict the domains formed by the dynamical quantities of eccentric non-orbital BBH mergers relative to circular BBH mergers. Sec. \ref{sec:IV} investigates whether similar domains exist in scenarios involving spin alignment and spin precession formed by effect of eccentricity. Finally, in Sec. \ref{sec:V}, we offer a comprehensive summary and outlook.
Throughout this work, we adhere to geometric units where $G=c=1$. The component masses of the BBH system are denoted as $m_1$ and $m_2$, with the total mass represented as $M$. The mass ratio $q$ is defined as $q=m_1 / m_2$, with $m_1$ being smaller than $m_2$.

\section{Methons}\label{sec:II}
\subsection{NR data}\label{sec:II:A}
This section provides an overview of the origin and composition of the NR simulation data concerning BBH mergers employed in our study. The data utilized are sourced from the SXS \cite{SXSBBH} and RIT \cite{RITBBH} NR catalogs. SXS employs a multi-domain spectral method \cite{Szilagyi:2009qz,Kidder:1999fv} incorporating a first-order iteration of the generalized harmonic formulation \cite{Hemberger:2012jz,Pretorius:2004jg,Garfinkle:2001ni} of Einstein’s equations with constraint damping for initial data evolution, alongside the Spectral Einstein Code (SpEC) \cite{SXSBBH} for conducting simulations. On the other hand, the simulations within the RIT catalog were progressed using the LazEv code \cite{Zlochower:2005bj} implementing the moving puncture approach \cite{Campanelli:2005dd}, in conjunction with the BSSNOK (Baumgarte-Shapiro-Shibata-Nakamura-Oohara-Kojima) formalism for evolution systems \cite{Marronetti:2007wz}. The LazEv code interfaces with the Cactus /Carpet /EinsteinToolkit infrastructure \cite{Lousto:2007rj,Zlochower:2012fk,Loffler:2011ay}.

While SXS primarily focuses on circular orbit simulations, RIT predominantly specializes in eccentric orbit simulations. The 4th release of the RIT catalog encompasses 824 eccentric BBH NR simulations, encompassing nonspinning, spin-aligned, and spin-precessing configurations with eccentricities spanning from 0 to 1 \cite{Healy:2022wdn}. We select a subset of nonspinning circular orbital simulations from SXS and RIT, as well as nonspinning simulations of eccentric orbits from RIT as the primary research dataset.

\begin{figure}[htbp!]
\centering
\includegraphics[width=8cm,height=5cm]{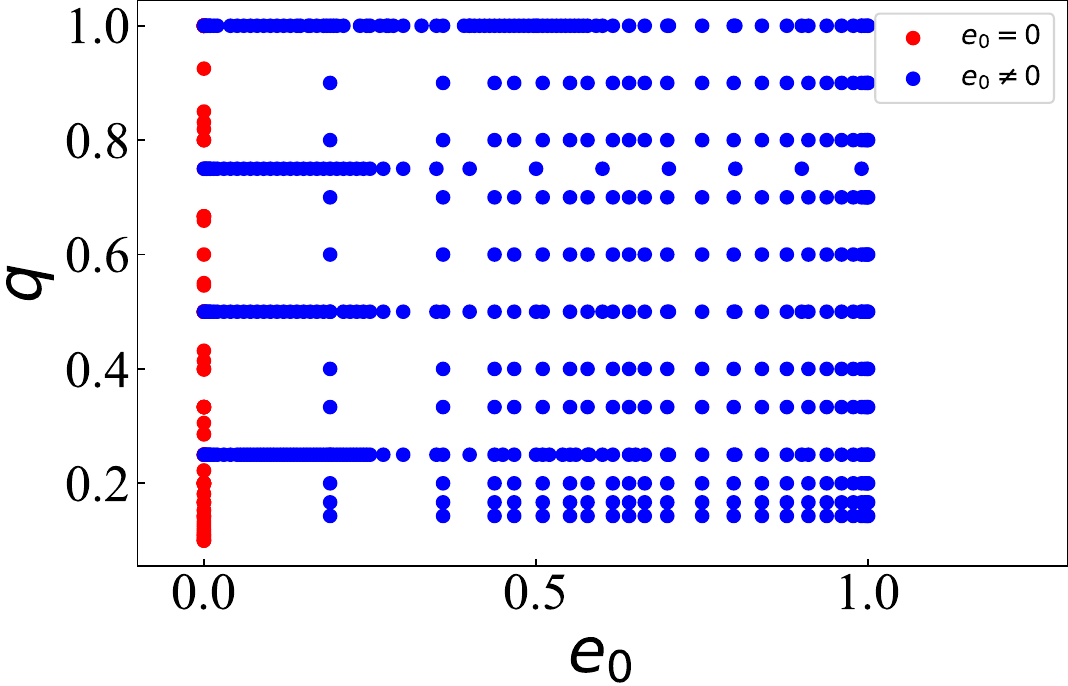}
\caption{\label{FIG:1}Parameter space of all nonspinning eccentric ($e_0\neq0$) orbital and circular ($e_0=0$) orbital simulations from SXS and RIT, which contains 80 sets of SXS circular orbital simulations, 21 sets of RIT circular orbital simulations, and 492 sets of RIT eccentric orbit simulations. The mass ratio $q$ of the circular orbital simulation is from 1/10 to 1, the mass ratio of the eccentric orbital simulation is from 1/7 to 1. The initial eccentricity $e_0$ is from 0 to 1.}
\end {figure}

FIG. \ref{FIG:1} illustrates the parameter space encompassing all eccentric orbital and circular orbital simulations we used. The initial eccentricity $e_0$ is derived from RIT's empirical formula $e_0=2 \epsilon-\epsilon^2$, where $\epsilon$ ranges from 0 to 1. Within FIG. \ref{FIG:1}, there are 80 sets of SXS circular orbit simulations, 21 sets of RIT circular orbit simulations, and 492 sets of eccentric orbit simulations. The mass ratio $q$ for the circular orbit simulations ranges from 1/10 to 1, while for the eccentric orbit simulations, it spans from 1/7 to 1. The initial eccentricity $e_0$ varies from 0 to 1. Notably, the RIT eccentric orbit simulations include two fixed initial distances of $11.3M$ and $24.6M$.

For the computation of dynamical quantities such as the peak luminosity $L_{\text{peak}}$, remnant mass $M_{\text{rem}}$, spin $\alpha_{\text{rem}}$, and recoil velocity $V_{\text{rem}}$, both SXS and RIT follow similar methodologies. RIT employs AHFinderDirect \cite{Thornburg:2003sf} to pinpoint apparent horizons and determines the horizon spins $S_H$ as the remnant spin $\alpha_{\text{rem}}$ through the isolated horizon algorithm. Subsequently, they calculate the horizon mass $m_\text{H}$ as the remnant mass $M_{\text{rem}}$ using the Christodoulou formula: $ m_\text{H} =\sqrt{m_{\mathrm{irr}}^2+S_\text{H}^2 /\left(4 m_{\mathrm{irr}}^2\right)}$, where $m_{\mathrm{irr}} = \sqrt{A_\text{H} /(16 \pi)}$ represents the irreducible mass defined by $A_\text{H}$, the surface area of the horizon \cite{Campanelli:2006fy}. Unlike mass $M_{\text{rem}}$ and spin $\alpha_{\text{rem}}$, recoil velocity $V_{\text{rem}}$ and peak luminosity $L_{\text{peak}}$ are derived from radiative quantities, i.e., the waveform. 

RIT provides waveform data in the form of the Newman-Penrose scalar $\Psi_4$ and gravitational wave strain $h$, which are downloadable from their catalog \cite{RITBBH}. These quantities can be expanded using the spin-weighted spherical harmonic function ${ }{-2} Y_{\ell , m}(\theta, \phi)$ with a spin weight $s=-2$. Specifically, the expansions are as follows:
\begin{equation}\label{eq:1}
r \Psi_4=\sum_{\ell, m} r \Psi_4^{\ell m}{ }_{-2} Y_{\ell, m}(\theta, \phi),
\end{equation}
and
\begin{equation}\label{eq:2}
r h=\sum_{\ell, m} r h^{\ell m}{}_{-2} Y_{\ell, m}(\theta, \phi),
\end{equation}
where $r$ denotes the extraction radius, and $h^{\ell m}$ and $\Psi_4^{\ell m}$ correspond to higher harmonic modes for $h$ and $\Psi_4$, respectively. As $r$ approaches infinity, the relationship $\Psi_4=\partial^2 h / \partial t^2$ holds. As detailed in Refs. \cite{Ruiz:2007yx}, the radiative quantities can be computed using $h$ and $\Psi_4$, where we take $h$ as an example.
The radiated energy $E_{\mathrm{rad}}$ can be determined as:
\begin{equation}\label{eq:3}
E_{\mathrm{rad}}=\lim _{r \rightarrow \infty} \frac{r^2}{16 \pi} \sum_{\ell, m} \int_{t_0}^t \mathrm{~d} t^{\prime}\left|\dot{h}^{\ell m}\right|^2,
\end{equation}
where $\dot{h}^{\ell m}$ represents the time derivative of $h^{\ell m}$. Subsequently, the peak luminosity $L_{\text {peak }}$ is calculated as the maximum of $\mathrm{d}E_{\mathrm{rad}}/\mathrm{d}t$:
\begin{equation}\label{eq:4}
L_{\text {peak }}=\max _t \mathrm{d}E_{\mathrm{rad}}/\mathrm{d}t.
\end{equation}
In this study, we focus solely on the nonspinning case, where the recoil velocity $V_{\mathrm{rem}}$ lies in the orbital plane, perpendicular to the orbital angular momentum $L$. It can be calculated as $V_{\mathrm{rem}}=\left|P_{\mathrm{rad}}\right| / M_{\mathrm{rem}}$, with:
\begin{equation}\label{eq:5}
\begin{aligned}
P_{\mathrm{rad}} & =\lim _{r \rightarrow \infty} \frac{r^2}{8 \pi} \sum_{\ell, m} \int_{t_0}^t \mathrm{~d} t^{\prime} \dot{h}^{\ell m} 
\left(a_{\ell, m} \bar{\dot{h}}^{\ell, m+1} \right. \\
&\left. +b_{\ell,-m} \bar{\dot{h}}^{\ell-1, m+1} -b_{\ell+1, m+1} \bar{\dot{h}}^{\ell+1, m+1}\right),
\end{aligned}
\end{equation}
where the coefficients $a_{\ell, m}$, $b_{\ell,-m}$, and $b_{\ell+1, m+1}$ can be referenced from Ref. \cite{Ruiz:2007yx}, and $\bar{\dot{h}}^{\ell m}$ represents the complex conjugate of $\dot{h}^{\ell m}$. Subsequently, we derive the values of both the peak luminosity $L_{\text {peak }}$ and the recoil velocity $V_{\mathrm{rem}}$ from $h$.

\subsection{PN method}\label{sec:II:B}
In this section, we introduce the 3PN order post-Newtonian nonspinning BBH inspiral waveform utilized in this investigation. As detailed in Refs. \cite{Hinder:2008kv,Memmesheimer:2004cv}, the 3PN conservative dynamics in harmonic coordinates can be parameterized by the PN expansion parameters $x \equiv(M \omega)^{2 / 3}$ and the temporal eccentricity $e_t$, where $\omega$ denotes the average orbital frequency. The PN expansion for the relative orbit radius $r$, angular frequency $\dot{\phi}$, mean anomaly $l$, and mean motion $n$ for nonspinning BBH is expressed as follows \cite{Hinder:2008kv}:
\begin{equation}\label{eq:6}
\begin{aligned}
r= & r_{\mathrm{Newt}} x^{-1}+r_{1 \mathrm{PN}}+r_{2 \mathrm{PN}} x
+r_{3 \mathrm{PN}} x^2 \\ 
& +\mathcal{O}\left(x^3\right)
\end{aligned},
\end{equation}
\begin{equation}\label{eq:7}
\begin{aligned}
\dot{\phi} = & \dot{\phi}_{ \mathrm{Newt}} x^{3 / 2}+\dot{\phi}_{1 \mathrm{PN}} x^{5 / 2} +\dot{\phi}_{2 \mathrm{PN}} x^{7 / 2} \\
&+\dot{\phi}_{3 \mathrm{PN}} x^{9 / 2}+\mathcal{O}\left(x^{11 / 2}\right)
\end{aligned},
\end{equation}
\begin{equation}\label{eq:8}
l=u-e_t \sin u+l_{2 \mathrm{PN}} x^2+l_{3 \mathrm{PN}} x^3+\mathcal{O}\left(x^4\right)
\end{equation}
and
\begin{equation}\label{eq:9}
\begin{aligned}
\dot{l} & =n \\
& =x^{3 / 2}+n_{1 \mathrm{PN}} x^{5 / 2}+n_{2 \mathrm{PN}} x^{7 / 2}+n_{3 \mathrm{PN}} x^{9 / 2}\\
&+\mathcal{O}\left(x^{11 / 2}\right)
\end{aligned},
\end{equation}
where $u$ signifies the eccentric anomaly, and $r_{1 \mathrm{PN}}$, $\dot{\phi}_{1 \mathrm{PN}}$, $l_{1 \mathrm{PN}}$, $n_{1 \mathrm{PN}}$, etc., represent the PN expansion coefficients, detailed in Ref. \cite{Hinder:2008kv}.

In the context of a real BBH system, considering only conservative dynamics is insufficient, as gravitational waves carry away energy and angular momentum, leading to the decay of the orbital distance and an increase in the angular frequency. Consequently, variations occur in $x$ and $e_t$. We incorporate the 3PN radiative dynamics from Refs. \cite{Arun:2009mc,Arun:2007rg,Arun:2007sg}, encompassing instantaneous and hereditary terms, expressed as:
\begin{equation}\label{eq:10}
\dot{x}= \dot{x}_{\mathrm{inst}}+\dot{x}_{\mathrm{hered}}
\end{equation}
and
\begin{equation}\label{eq:11}
\dot{e_t}= \dot{e_t}_{\mathrm{inst}}+\dot{e_t}_{\mathrm{hered}},
\end{equation}
where the subscripts \textit{inst} and \textit{hered} mean \textit{instantaneous} and \textit{hereditary}. Their specific expressions are
\begin{equation}\label{eq:12}
\begin{aligned}
\dot{x}_{\mathrm{inst}}= & \frac{2 \eta}{3  M} x^5 \left(\dot{x}_{\mathrm{Newt}}+\dot{x}_{1 \mathrm{PN}} x+\dot{x}_{2 \mathrm{PN}} x^2 \right. \\
& \left. +\dot{x}_{3 \mathrm{PN}} x^3\right)
\end{aligned},
\end{equation}

\begin{equation}\label{eq:13}
\begin{aligned}
\dot{e_t}_{\mathrm{inst}}= & -\frac{ \eta}{ M} e_t x^4 \left(\dot{e_t}_{\mathrm{Newt}}+\dot{e_t}_{1 \mathrm{PN}} x+\dot{e_t}_{2 \mathrm{PN}} x^2 \right. \\
& \left. +\dot{e_t}_{3 \mathrm{PN}} x^3 \right)
\end{aligned},
\end{equation}

\begin{equation}\label{eq:14}
\begin{aligned}
\dot{x}_{\mathrm{hered}}= & \frac{64 \eta}{5  M} x^4 \left(\dot{x}_{1.5\mathrm{PN}} x^{3/2}+\dot{x}_{2.5 \mathrm{PN}} x^{5/2} \right. \\
&\left. +\dot{x}_{3 \mathrm{PN}} x^3 \right)
\end{aligned}
\end{equation}
and 

\begin{equation}\label{eq:15}
\begin{aligned}
\dot{e_t}_{\mathrm{hered}}= & \frac{32 \eta}{5 M} e_t x^4 \left(\dot{e_t}_{1.5\mathrm{PN}} x^{3/2}+\dot{e_t}_{2.5 \mathrm{PN}} x^{5/2} \right. \\
&\left. +\dot{e_t}_{3 \mathrm{PN}} x^3 \right)
\end{aligned},
\end{equation}
where $\dot{x}_{1 \mathrm{PN}}$, $\dot{e_t}_{1 \mathrm{PN}}$, etc., denote the PN expansion coefficients, which can be found in Ref. \cite{Arun:2009mc}.

The radiative dynamics described by Eqs. (\ref{eq:10}) and (\ref{eq:11}) constitute a self-contained system, solvable independently of the aforementioned conservative dynamics once the initial conditions ${e_t}_0$ and $x_0$ are provided.
Upon obtaining solutions for $e_t$ and $x$, we integrate Eq. (\ref{eq:9}) to determine $l(t)$, considering an initial mean anomaly $l_0$ at $t_0$. Subsequently, by substituting $l(t)$ into Eq. (\ref{eq:8}) and numerically solving for the eccentric anomaly $u$ using root-finding techniques, we then substitute $u$, $e_t$, and $x$ into Eqs. (\ref{eq:6}) and (\ref{eq:7}) to derive $r$ and $\dot{\phi}$. Further differentiation and integration of these quantities yield $\dot{r}$ and $\phi$, respectively. Following this process, we successfully resolve the dynamics governing the inspiral phase of BBH in eccentric orbits.

For computing the gravitational wave waveform strain $h$, we employ the 3PN nonspinning instantaneous waveform from Ref. \cite{Mishra:2015bqa}, with its harmonic modes expressed as:
\begin{equation}\label{eq:16}
h^{\ell m}=\frac{4  M \eta}{ R} \sqrt{\frac{\pi}{5}} e^{-i m \phi} H^{\ell m},
\end{equation}
where $R$ is the observation distance, which can be set to 1 during calculation. Each mode corresponds to a distinct PN order and cannot be simultaneously expressed. Taking the $h^{22}$ mode as an illustration, it can be represented as:
\begin{equation}\label{eq:17}
h^{22}=\frac{4  M \eta }{ R} \sqrt{\frac{\pi}{5}} e^{-2 i \phi} H^{22},
\end{equation}
where the amplitude $H^{22}$ is a composite of terms from various PN orders:
\begin{equation}\label{eq:18}
H^{22}=H_{\text {Newt }}^{22}+H_{1 \mathrm{PN}}^{22}+H_{2 \mathrm{PN}}^{22}+H_{2.5 \mathrm{PN}}^{22}+H_{3 \mathrm{PN}}^{22}.
\end{equation}
The leading Newtonian order term in Eq. (\ref{eq:18}) corresponds to the well-known quadrupole moment formula frequently utilized. Detailed formulas for the amplitudes in Eq. (\ref{eq:18}) and other modes in Eq. (\ref{eq:16}) can be referenced in Ref. \cite{Mishra:2015bqa}.

\subsection{Envelope of the radiative quantities}\label{sec:II:C}
In our previous work \cite{Wang:2023vka}, we observed oscillations in dynamic quantities peak luminosity $L_{\text{peak}}$, mass $M_{\text{rem}}$, spin $\alpha_{\text{rem}}$, and recoil velocity $V_{\text{rem}}$ concerning changes in the initial eccentricity $e_0$. These oscillations were effectively elucidated by associating them with integer values of the gravitational wave cycle.

In another study \cite{Wang:2024jro}, we conducted a complete comparison between NR and PN waveforms of nonspinning and spin-aligned BBHs in eccentric orbits, focusing on waveforms with a duration of start to $200M$ before the merger, representing the inspiral phase. Employing the least squares method to fit the frequency of $\Psi_4^{22}$, we precisely determined the initial parameters ${e_t}_0$, $x_0$, and $l_0$ for all eccentric waveforms that could be effectively aligned with NR simulations conducted by SXS and RIT.

Furthermore, in a separate investigation \cite{Wang:2024afj}, we reexamined the oscillatory behavior previously identified \cite{Wang:2023vka}. This time, we segmented the RIT's eccentric orbital waveform into four distinct phases: inspiral (200M before the merger), near-merger (200M to the moment of merger), merger moment (at maximum amplitude), and ringdown (merger to the final stage). Surprisingly, we discovered that all four phases exhibited a consistent oscillation pattern. By comparing the radiated energy from eccentric PN waveforms and its orbital average energy flux to NR waveforms, we determined that the oscillations stemmed from non-orbital averaging, specifically influenced by the initial mean anomaly $l_0$. For a fixed initial eccentricity $e_0$ and PN expansion parameters $x_0$, varying $l_0$ results in different radiated energies. Notably, if orbital averaging is applied, the averaged radiation energy will become independent of $l$. These facts suggest that it is a unique and universal effect of eccentricity across all phases of eccentric binary black hole mergers. This effect stands in stark contrast to circular orbit mergers. Our findings also revealed a direct relationship: the higher the eccentricity, the more pronounced the oscillations, explaining the escalating nature of oscillations observed in our previous work \cite{Wang:2023vka}.

Here, we present the method utilized to compute the radiated angular momentum in this investigation. As detailed in Ref. \cite{Ruiz:2007yx}, owing to symmetry considerations, the radiated angular momentum $L^{\mathrm{rad}} = L_z^{\mathrm{rad}}$ predominantly aligns with the $z$ direction, coinciding with the orbital angular momentum direction $L$. This quantity is determined by:
\begin{equation}\label{eq:19}
L_{z}^{\mathrm{rad}} =\lim _{r \rightarrow \infty} \frac{ r^2}{16 \pi} \operatorname{Im}\left\{\sum_{\ell, m} m \int_{t_0}^t \left( h^{\ell m} \bar{\dot{h}}^{\ell m} \right)\mathrm{~d} t^{\prime}\right\},
\end{equation}
where $\operatorname{Im}$ means taking the imaginary part.

The initial mean anomaly $l_0$ in both NR and PN waveforms discussed in Refs. \cite{Wang:2024jro} and \cite{Wang:2024afj} remains constant. While we are unable to alter the $l_0$ in NR waveforms, we have the flexibility to adjust the $l_0$ in PN waveforms that are fitted to NR waveforms during the inspiral phase. This adjustment allows us to completely compute dynamic quantities such as radiated energy $E_{\mathrm{rad}}$, radiated angular momentum $L_{\mathrm{rad}}$, and radiated linear momentum $P_{\mathrm{rad}}$ using Eqs. (\ref{eq:3}), (\ref{eq:5}), and (\ref{eq:19}) for all possible $l_0$. The initial PN parameters $l_0$, ${e_t}_0$, and $x_0$, along with the PN waveform models for evaluating these dynamic quantities, can be found in Ref. \cite{Wang:2024jro} and Sec. \ref{sec:II:B}.
The eccentric orbit waveforms from RIT correspond to specific initial $l_0$ values. However, by manipulating $l_0$, the resultant calculated radiative quantities correspond to distinct values, preserving the original ${e_t}_0$ and $x_0$ parameters but with a different $l_0$. As we systematically vary $l_0$ across the range from 0 to $2\pi$, all feasible values of the radiative quantities are explored, encapsulating a specific range.

\begin{figure*}[htbp!]
\centering
\includegraphics[width=16cm,height=5cm]{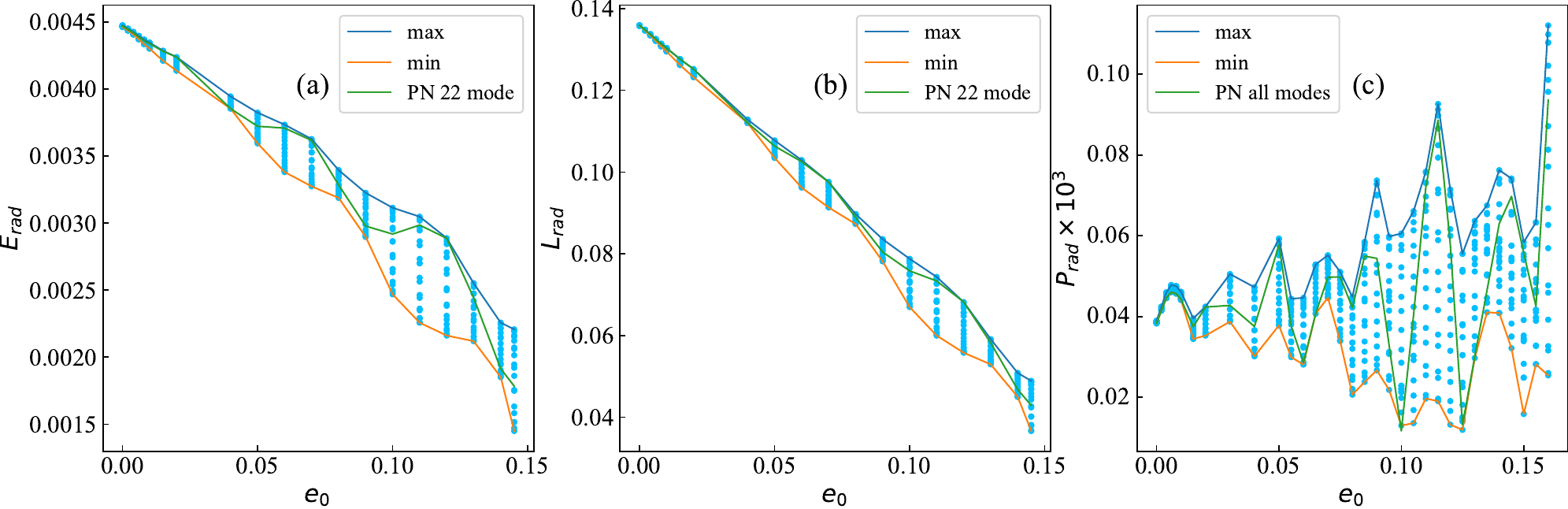}
\caption{\label{FIG:2}Values and ranges of radiative quantities—radiated energy $E_{\mathrm{rad}}$, radiated angular momentum $L_{\mathrm{rad}}$, and radiated linear momentum $P_{\mathrm{rad}}$, which are calculated from PN waveforms fitted to NR waveforms during the inspiral phase (200M before the merger) for specific mass ratios, as the initial mean anomaly $l_0$ varies continuously from 0 to $2\pi$. panels (a) and (b) correspond to radiated energy and angular momentum calculated with mass ratio $q=1$ by 22 modes, and panel (c) corresponds to radiated linear momentum calculated with mass ratio $q=1/2$ by all modes.}
\end {figure*}

FIG. \ref{FIG:2} illustrates the values and ranges of radiative quantities—radiated energy $E_{\mathrm{rad}}$, radiated angular momentum $L_{\mathrm{rad}}$, and radiated linear momentum $P_{\mathrm{rad}}$, which are calculated from PN waveforms fitted to NR waveforms during the inspiral phase (200M before the merger) for specific mass ratios, as the initial mean anomaly $l_0$ varies continuously from 0 to $2\pi$. The initial eccentricity $e_0$ from RIT is utilized here for consistency with previous works \cite{Wang:2024afj}. The utilization of ${e_t}_0$ is also permissible in this context, as its values closely resemble those of $e_0$ and can effectively represent the measured eccentricity.

In panels (a) and (b), we present the relationship between radiated energy $E_{\mathrm{rad}}$ and radiated angular momentum $L_{\mathrm{rad}}$ calculated using PN waveforms with a mass ratio of $q=1$, and the initial eccentricity $e_0$. The green line representing the PN 22 mode in the figure showcases the oscillatory trend (originally shown as scatter points). The light blue scatter points depict the results obtained by adjusting the $l_0$ value continuously around each original PN fitting scatter point within the range [0, $2\pi$] for 20 consecutive points. The maximum and minimum lines denote the extreme values among these scatter points. Our calculations in panels (a) and (b) solely consider the 22 mode, given its primary contribution to radiated energy and angular momentum.

In panel (c), we depict the relationship between radiated linear momentum $P_{\mathrm{rad}}$, computed using PN waveforms with a mass ratio of $q=1/2$, and the initial eccentricity $e_0$. The representation of lines and points in panel (c) mirrors that of panels (a) and (b). Notably, we employ a mass ratio of $q=1/2$ here, as the radiated linear momentum is zero when $q=1$ due to symmetry considerations. Furthermore, our calculation results incorporate $\ell \leq 4$ in Eq. (\ref{eq:5}), as higher-order modes significantly impact the results, as indicated in Ref. \cite{Radia:2021hjs}.

In FIG. \ref{FIG:2}, each vertical points corresponds to a specific initial eccentricity $e_0$. However, as $l_0$ continuously changes, these points progressively populate the space between the maximum and minimum bounds. This observation underscores that the waveforms fitted in previous works \cite{Wang:2024afj} (PN 22 mode and PN all modes in FIG. \ref{FIG:2}) represent merely a subset of $l_0$ values within the outlined range, and the oscillation of radiated energy with eccentricity is a particular manifestation of this interplay. Notably, the continuous variation of $l_0$ is primarily aimed at maximizing the exploration of the parameter space $[0,2\pi]$ and does not carry a specific significance.

Contrary to a scenario where adjusting $l_0$ might lead to a linear progression of $E_{\mathrm{rad}}$, $L_{\mathrm{rad}}$, and $P_{\mathrm{rad}}$ from a minimum to maximum at a given eccentricity, our findings demonstrate oscillations in these dynamic quantities. While the initial eccentricities displayed in FIG. \ref{FIG:2} may not be densely distributed, resulting in visible gaps between different eccentricities, it is foreseeable that denser eccentricity distributions would obscure these gaps, leaving behind regions delineated by the maximum and minimum values.
These regions effectively serve as envelopes encapsulating all variations stemming from alterations in $l_0$ and $e_0$, hence why this section is aptly named the `envelope of the radiative quantities.' The envelopes depicted in FIG. \ref{FIG:2} exhibit similarities and distinctions. Notably, as eccentricity rises, the envelope's span widens, signifying a heightened effect of eccentricity on these quantities. Both radiative energy and angular momentum envelopes display a consistent increase, while the radiative linear momentum envelope showcases irregularities with a larger range compared to the energy and angular momentum envelopes; furthermore, the energy envelope surpasses that of the angular momentum.

\begin{figure*}[htbp!]
\centering
\includegraphics[width=16cm,height=16cm]{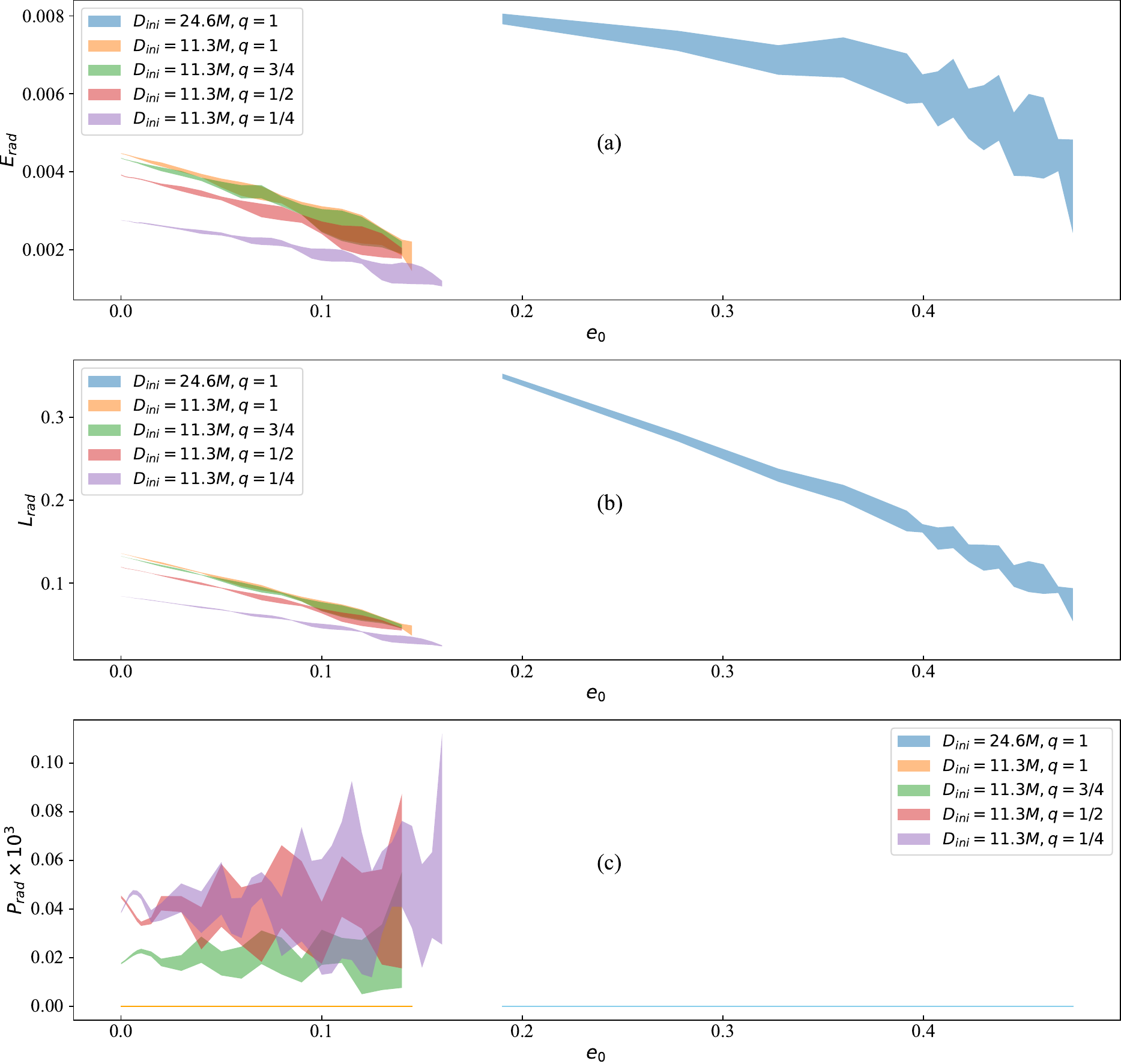}
\caption{\label{FIG:3}Envelopes of radiative quantities delineated by these five cases ($D_{\text{ini}}=24.6 M$ with $q=1$, $D_{\text{ini}}=11.3 M$ with $q=1$, $D_{\text{ini}}=11.3 M$ with $q=3/4$, $D_{\text{ini}}=11.3 M$ with $q=1/2$, and $D_{\text{ini}}=11.3 M$ with $q=1/4$) with the initial eccentricity for varying $l_0$ in $[0,2\pi]$.}
\end {figure*}

The emergence of oscillations in the radiative quantities displayed in FIG. \ref{FIG:2} hinges on the presence of NR simulations with a fixed initial distance and a sufficiently broad array of continuously varying initial eccentricities. The oscillations in radiative quantities are fundamentally driven by the continuous adjustment of initial eccentricities. In scenarios where the number of NR simulations is limited, such oscillatory patterns may remain imperceptible. RIT has conducted a series of BBH merger simulations involving eccentric orbits, with two distinct initial distances: $D_{\text{ini}}=11.3 M$ and $D_{\text{ini}}=24.6 M$. Among these simulations, only a subset featuring continuously changing eccentricities in five cases vividly exhibit the oscillatory behavior of dynamic quantities. In contrast, other cases lack observable oscillations due to the limited number of simulations.
The five cases corresponding to specific initial distances and mass ratios are as follows: $D_{\text{ini}}=24.6 M$ with $q=1$, $D_{\text{ini}}=11.3 M$ with $q=1$, $D_{\text{ini}}=11.3 M$ with $q=3/4$, $D_{\text{ini}}=11.3 M$ with $q=1/2$, and $D_{\text{ini}}=11.3 M$ with $q=1/4$. FIG. \ref{FIG:3} illustrates the envelopes of radiative quantities delineated by these five cases with the initial eccentricity for varying $l_0$ in $[0,2\pi]$, reflecting the characteristics elucidated in FIG. \ref{FIG:2}. Notably, in the case of $D_{\text{ini}}=24.6 M$, the prolonged inspiral phase of the BBHs results in longer waveforms and larger radiative quantities compared to the scenario with $D_{\text{ini}}=11.3 M$. Given that linear momentum radiation is null for mass ratio of $q=1$, it is represented solely as a line in panel (c) of FIG. \ref{FIG:3}.

As the radiative quantities form envelopes, the dynamic quantities $\alpha_{\text{rem}}$, $M_{\text{rem}}$, and $V_{\text{rem}}$ also encapsulate similar envelope-like structures. The observed oscillations highlighted in Refs. \cite{Wang:2023vka,Wang:2024afj} are therefore special cases arising from special initial conditions.

\subsection{Modeling dynamical quantities in circular orbits}\label{sec:II:D}
Before delving into the impact of eccentricity on modeling the dynamical quantities $M_{\text{rem}}$, $\alpha_{\text{rem}}$, $V_{\text{rem}}$, and $L_{\text{peak}}$ in BBH mergers, it is important to revisit how dynamical quantities are modeled in circular orbits. Notably, we incorporate the peak luminosity $L_{\text{peak}}$, given its substantive importance, as eccentricity exerts a comparable influence on both the inspiral phase and the moment of merger. The modeling process typically involves establishing a connection between initial parameters like mass ratios, spins of constituent black holes, etc., and the final quantities through a defined methodology. In scenarios involving the merger of nonspinning BBHs in circular orbits, the initial separation marginally impacts the eventual dynamical quantities. Consequently, these quantities are predominantly contingent on the mass ratio $q$. We employ a straightforward modeling approach—polynomial fitting, akin to the methodology employed that we introduced in the introduction.

\begin{figure*}[htbp!]
\centering
\includegraphics[width=16cm,height=8cm]{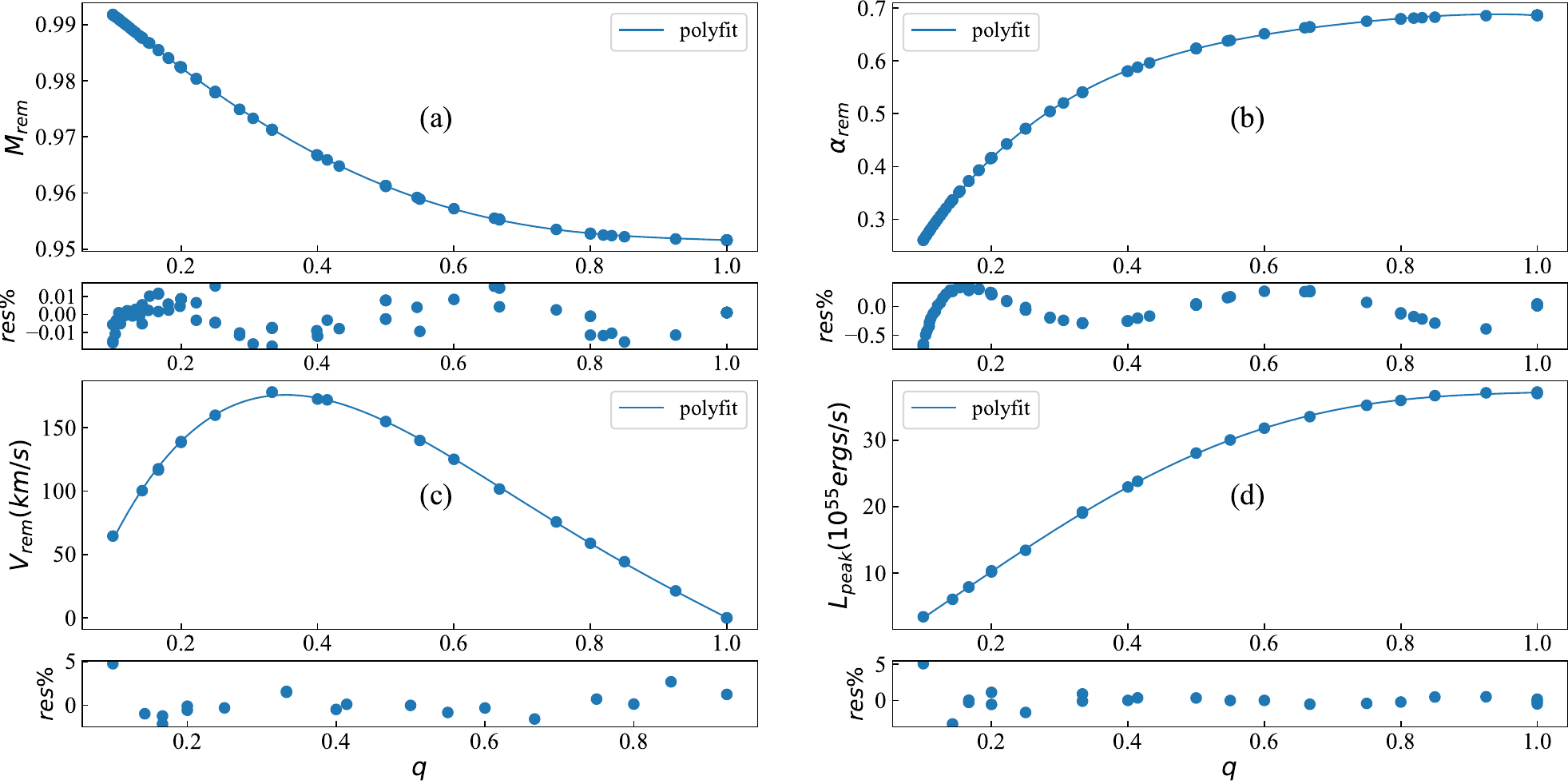}
\caption{\label{FIG:4}Polynomial modeling and residuals for the dynamical quantities $M_{\text{rem}}$, $\alpha_{\text{rem}}$, $V_{\text{rem}}$, and $L_{\text{peak}}$ of BBH mergers in circular orbits with respect to the mass ratio $q$.}
\end {figure*}

In FIG. \ref{FIG:4}, we conduct polynomial modeling for the dynamical quantities $M_{\text{rem}}$, $\alpha_{\text{rem}}$, $V_{\text{rem}}$, and $L_{\text{peak}}$ of BBH mergers in circular orbits with respect to the mass ratio $q$. The parameter space of FIG. \ref{FIG:4} is detailed in FIG. \ref{FIG:1}. We employ 4th order polynomials for all modeling endeavors. Panels (a) and (b) encompass 100 data points sourced from both the SXS and RIT catalogs, whereas panels (c) and (d) feature 22 data points exclusively from RIT with no pertinent data available from SXS. The variance in data points across the panels does not significantly impact the modeling accuracy. To evaluate the quality of our modeling, we calculate the percentage residual using the formula:
\begin{equation}\label{eq:20}
\text{res}=\frac{A_{\text{poly}}-A}{A} \times 100 \% ,  
\end{equation}
where $A$ represents $M_{\text{rem}}$, $\alpha_{\text{rem}}$, $V_{\text{rem}}$, and $L_{\text{peak}}$ respectively, and $A_{\text{poly}}$ denotes the polynomial fit of $A$. Reviewing FIG. \ref{FIG:4}, we observe that the percentage residuals from the polynomial modeling for the dynamical quantities $M_{\text{rem}}$, $\alpha_{\text{rem}}$, $V_{\text{rem}}$, and $L_{\text{peak}}$ fall within 0.02\%, 0.7\%, 5\%, and 5\% respectively.

\begin{figure*}[htbp!]
\centering
\includegraphics[width=16cm,height=10cm]{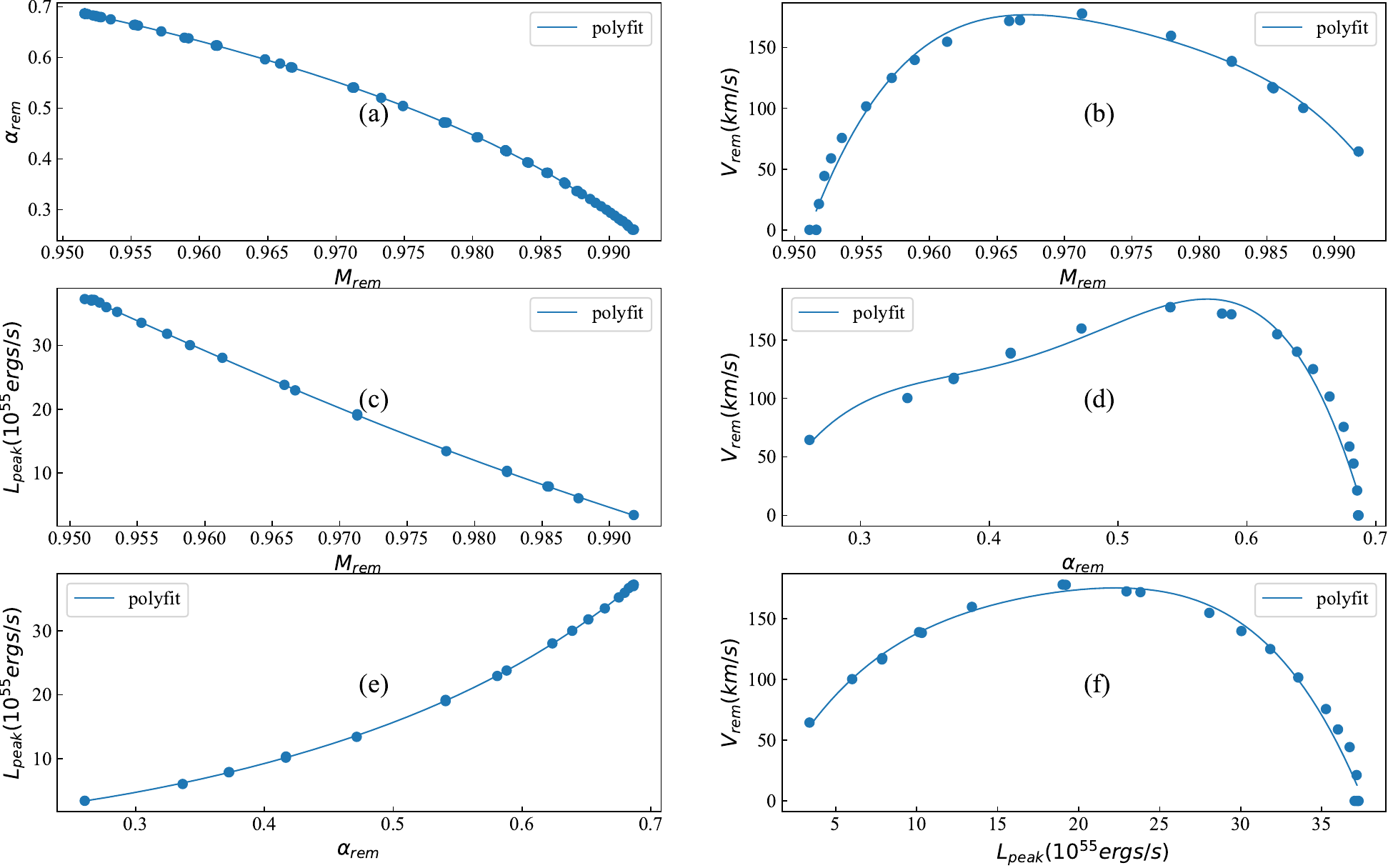}
\caption{\label{FIG:5}Polynomial modeling for the six correlations between dynamic quantities $M_{\text{rem}}$, $\alpha_{\text{rem}}$, $V_{\text{rem}}$, and $L_{\text{peak}}$ of BBH mergers in circular orbits.}
\end {figure*}

In addition to modeling the relationships between dynamical quantities and the mass ratio, we can also explore their correlations. Understanding these correlations are crucial for deciphering the strong-field dynamics of BBH mergers. Notably, for circular orbits, these quantities indeed exhibit intrinsic correlations. In FIG. \ref{FIG:5}, we exemplify the six correlations between these dynamic quantities using polynomial modeling, akin to FIG. \ref{FIG:4}. The percentage residuals among them are approximately within the same order of magnitude as the mass ratio modeling in FIG. \ref{FIG:4}; hence, specific calculations are omitted in this context. FIG. \ref{FIG:5} illustrates that polynomials effectively capture the correlations between these dynamic quantities.

\subsection{Combining eccentric and circular orbits}\label{sec:II:E}
In the preceding section, we employed polynomial fitting to explore the relationships between the dynamical quantities $M_{\text{rem}}$, $\alpha_{\text{rem}}$, $V_{\text{rem}}$, and $L_{\text{peak}}$ in circular orbits and the mass ratio, as well as their correlations. Our findings indicate that these relationships in circular orbits can be effectively modeled by polynomials.

In our prior work \cite{Wang:2023vka}, we proposed that the general trend regarding the impact of eccentricity on dynamical quantities $M_{\text{rem}}$, $\alpha_{\text{rem}}$, $V_{\text{rem}}$, and $L_{\text{peak}}$ is characterized by a gradual oscillation with varying initial eccentricity. Furthermore, we observed that the magnitude of oscillations intensifies with higher eccentricities. Once these quantities oscillate to their maximum or minimum values, constrained by the initial distance, the motion of the BBHs transitions from orbital merger (where the orbital cycle exceeds 1 or the gravitational wave phase spans more than $4\pi$ from the initial moment to the merger moment) to non-orbital merger (where the orbital cycle is less than 1 or the gravitational wave phase is less than $4\pi$). During non-orbital merger, the dynamical quantities progressively converge from peak values to stable values or zero as eccentricity varies.

\begin{figure*}[htbp!]
\centering
\includegraphics[width=16cm,height=20cm]{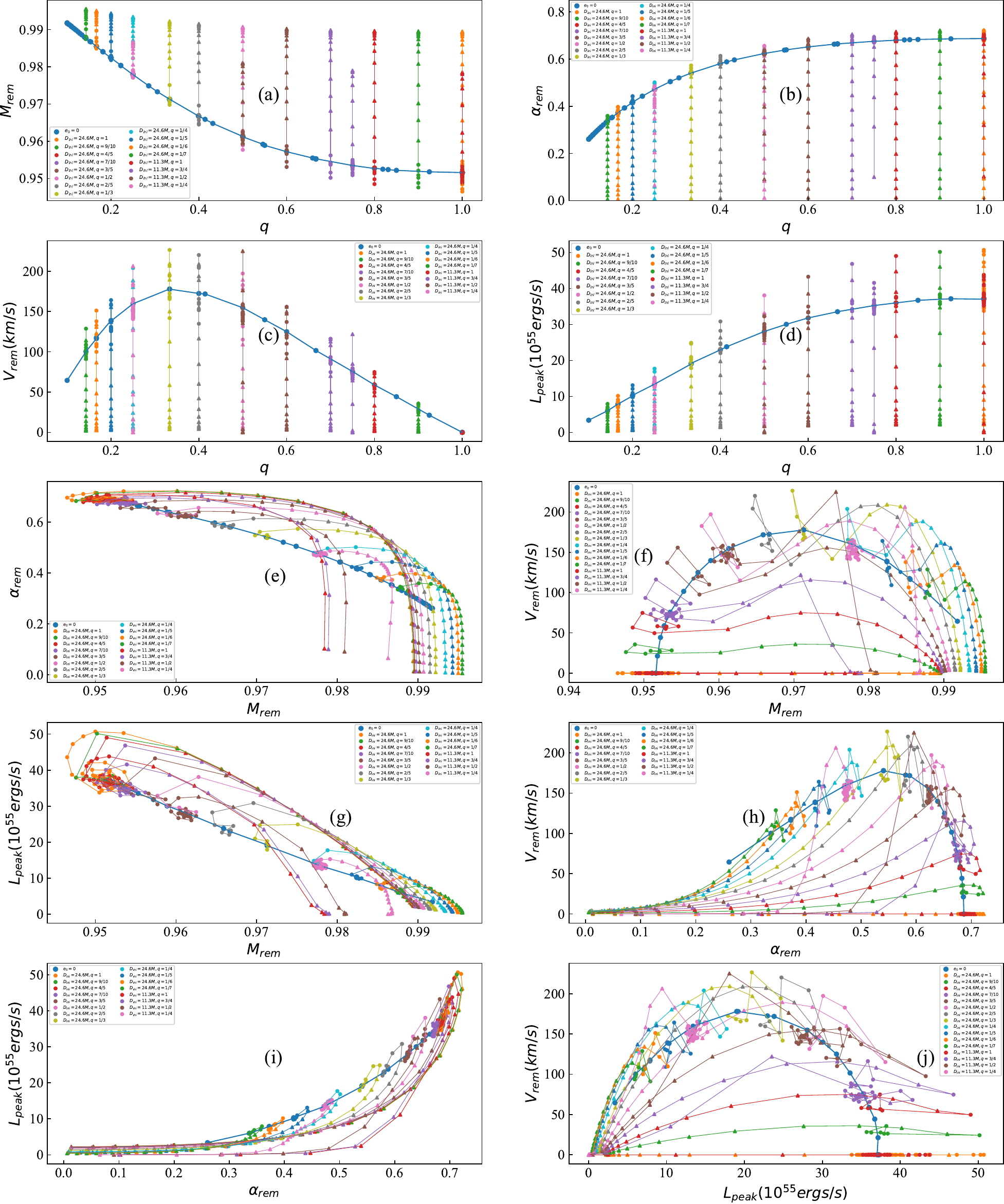}
\caption{\label{FIG:6}Summary of the relationships between the dynamical quantities $M_{\text{rem}}$, $\alpha_{\text{rem}}$, $V_{\text{rem}}$, and $L_{\text{peak}}$ and the mass ratio and their correlations, combining eccentric and circular orbits. Orbital mergers are represented by circular points, while non-orbital mergers are denoted by triangular points. The broken lines connect these points, only illustrating the changing trends of the dynamic quantities.}
\end {figure*}

In FIG. \ref{FIG:6}, we present a summary of the relationships between the dynamical quantities $M_{\text{rem}}$, $\alpha_{\text{rem}}$, $V_{\text{rem}}$, and $L_{\text{peak}}$ and the mass ratio and their correlations, combining eccentric and circular orbits. Orbital mergers are represented by circular points, while non-orbital mergers are denoted by triangular points. The broken lines in FIG. \ref{FIG:6} connect these points, only illustrating the changing trends of the dynamic quantities.
The data points in FIG. \ref{FIG:6} comprise information from circular orbits ($e_0=0$) as well as eccentric orbit ($e_0\neq0$) data with initial distances of $11.3M$ and $24.6M$, as outlined in the parameter space depicted in FIG. \ref{FIG:1}.

In panels (a), (b), (c), and (d) of FIG. \ref{FIG:6}, we present the relationships between dynamical quantities $M_{\text{rem}}$, $\alpha_{\text{rem}}$, $V_{\text{rem}}$, and $L_{\text{peak}}$ and the mass ratio $q$ within circular and eccentric orbits. It is evident that eccentricity plays a substantial role in both orbital and non-orbital BBH mergers.

For orbital BBH mergers, eccentricity's influence on the dynamical quantities remains constrained to a vicinity around the values of circular orbital BBH mergers. These quantities oscillate in the vicinity of the value of a circular orbit for a specific mass ratio as eccentricity varies. With sufficient NR data or densely initial eccentricities, the vicinity of dynamical quantities for a given mass ratio expands as the initial distance increases, as evident in cases with mass ratios of $11.3M$ and $24.6M$ where the eccentricities are dense enough. This effect intensifies with larger mass ratios, shaping the impact of mass ratio on eccentric orbital BBH mergers.

In contrast, for non-orbital mergers, eccentricity's influence spans a broader range relative to orbital mergers. This variation is manifested as the dynamical quantities transition from a value to a stable value or zero as BBHs progress from high eccentricities towards head-on collisions ($e_0=1$). The range of non-orbital mergers sometimes differs from that of orbital mergers; for instance, in panel (a), $M_{\text{rem}}$ showcases distinct ranges for orbital and non-orbital mergers. However, in panels (b), (c), and (d), the ranges of non-orbital mergers encompass those of orbital mergers in $\alpha_{\text{rem}}$, $V_{\text{rem}}$, and $L_{\text{peak}}$. This overlap results in identical dynamical quantities corresponding to different initial eccentricities or dynamics processes, leading to degeneracy.

In the scenario of $11.3M$, the dynamical quantities of non-orbital mergers fail to reach the head-on collision values observed at $24.6M$. This discrepancy arises from the insufficient density of initial eccentricities in the non-orbital merger case for $11.3M$, preventing the attainment of head-on collision values. This discrepancy is evident in the case of $11.3M$ and $q=0.75$ in panel (b); ideally, $\alpha_{\text{rem}}$ should reach 0 in a head-on collision scenario, which is not observed in this case.

In panels (e), (f), (g), (h), (i), and (j) of FIG. \ref{FIG:6}, we illustrate the correlation between the dynamical quantities $M_{\text{rem}}$, $\alpha_{\text{rem}}$, $V_{\text{rem}}$, and $L_{\text{peak}}$ in circular and eccentric orbits. In our previous work \cite{Wang:2023wol}, we highlighted that these dynamical quantities exhibit spiral or other complex structures in orbital mergers, while demonstrating regular correlations in non-orbital mergers, converging from specific values to head-on collision values. The core of the spiral or other structures corresponds to the dynamical quantities of the circular orbit with the respective mass ratio. Specifically, the relationship between $M_{\text{rem}}$, $\alpha_{\text{rem}}$, and $L_{\text{peak}}$ forms spiral patterns, whereas the association with $V_{\text{rem}}$ results in a complex, non-spiral structure due to the intricate behavior of $V_{\text{rem}}$ as it varies with eccentricity \cite{Wang:2023wol}. These descriptions align with panels (a), (b), (c), and (d), and detailed explanations are omitted here.

In essence, all these outcomes arise due to eccentricity causing variations in the relationships between the dynamical quantities, the mass ratio, and their correlations for both orbital and non-orbital BBH mergers compared to circular orbits.

\section{Results}\label{sec:III}
Based on the analysis in Sec. \ref{sec:II:C}, it is evident that the oscillation discussed in Ref. \cite{Wang:2023vka} represents a specific scenario stemming from constraints in initial conditions, while the genuine effect of eccentricity should be encapsulated within an envelope. The impact on the radiative quantities parallels the influence on the dynamical quantities $M_{\text{rem}}$, $\alpha_{\text{rem}}$, $V_{\text{rem}}$, and $L_{\text{peak}}$ as they are intricately connected—the dynamical quantities can be derived from the radiative quantities.

In FIG. \ref{FIG:6}, we present the relationships between dynamical quantities and mass ratio in eccentric orbits. Notably, eccentricity induces dynamical quantities to span a range through vertical scatter points relative to circular orbits. When the initial eccentricities from NR are sufficiently dense, these scatter points within these ranges transition from discrete points to continuous variations, whether for orbital or non-orbital mergers.

Simultaneously, in conjunction with the envelope formed by eccentricity in Sec. \ref{sec:II:B}, this phenomenon also leads to the densification and continuous alteration of the ranges formed by these scatter points. Nevertheless, considering the current limitations imposed by initial distances and the quantity of NR simulations, the ranges delineated by eccentricity in panels (a), (b), (c), and (d) of FIG. \ref{FIG:6} solely symbolize the maximum extent of the effects observed in NR studies thus far. The actual ranges are expected to expand as the initial distance increases. This expansion is attributed to the capacity of larger initial distances to accommodate heightened initial eccentricities during the transition from orbital to non-orbital mergers, correlating with the pronounced oscillations observed in Ref. \cite{Wang:2023vka} or the extended range of dynamical quantities for orbital mergers in FIG. \ref{FIG:6}.

Next, we will elucidate the impact of eccentricity on dynamic quantity modeling concerning both orbital and non-orbital merger.

\subsection{Orbital merger}\label{sec:III:A}
\begin{figure*}[htbp!]
\centering
\includegraphics[width=16cm,height=16cm]{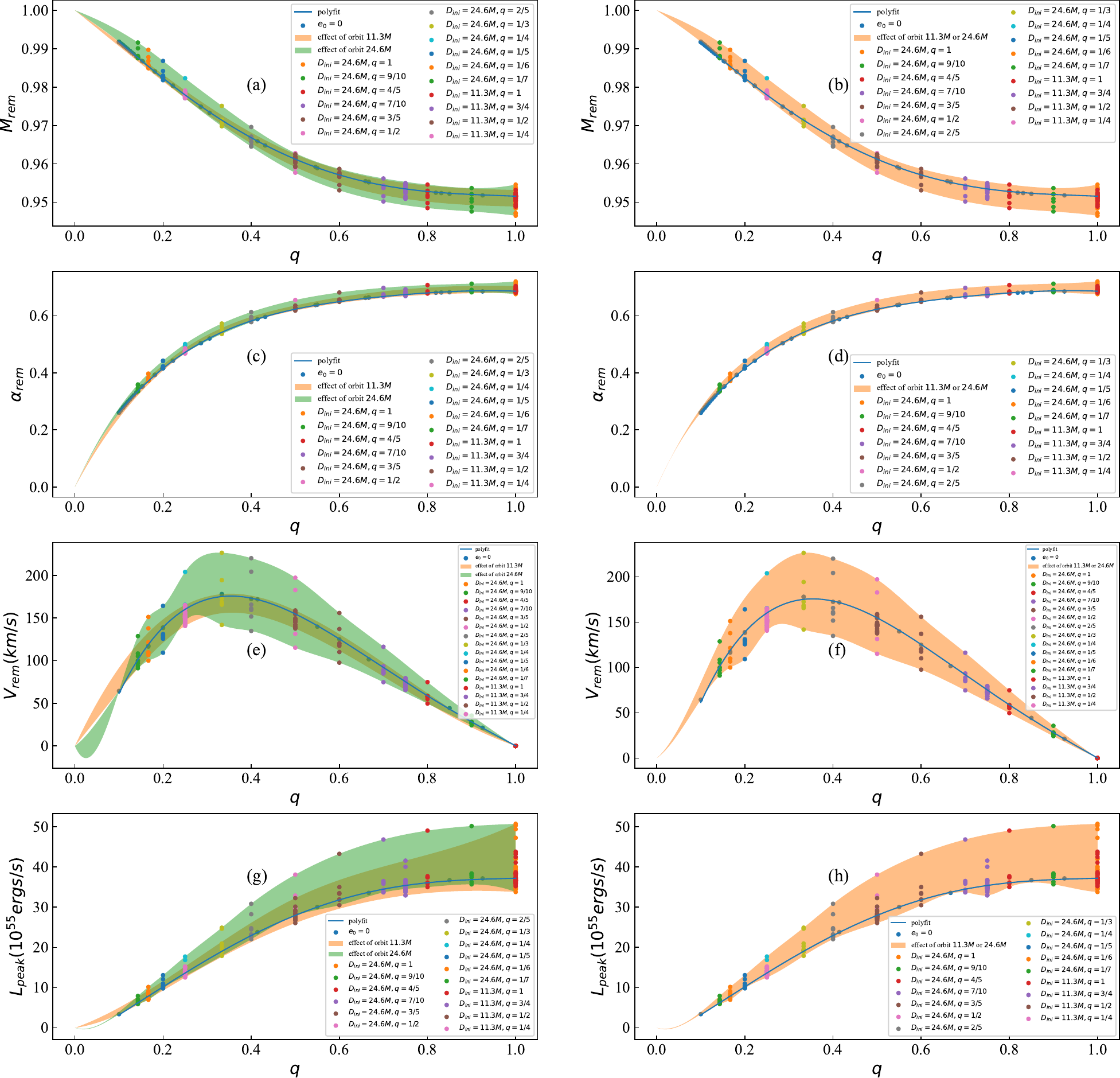}
\caption{\label{FIG:7}Domains shaped by the influence of eccentricity beyond the circular orbit on the dynamical quantities. In the left panels (a), (c), (e), and (g), we illustrate the domain synthesized through interpolation of the maximum and minimum values for two distinct initial distances of orbital BBH mergers: $11.3M$ and $24.6M$. In the right panels (b), (d), (f), and (h), we showcase the domains derived from the combination of data points for $11.3M$ and $24.6M$.}
\end {figure*}

In FIG. \ref{FIG:7}, we delineate the domains shaped by the influence of eccentricity beyond the circular orbit on the dynamical quantities.

In the left panels (a), (c), (e), and (g), we illustrate the domain synthesized through interpolation of the maximum and minimum values (i.e., boundaries) of the dynamical quantities concerning the mass ratio for two distinct initial distances of orbital BBH mergers: $11.3M$ and $24.6M$. During this interpolation process, we incorporate points where the dynamical quantities are established at the limit of $q=0$, reflected as $M_{\text{rem}}=1$, $\alpha_{\text{rem}}=0$, $V_{\text{rem}}=0$, and $L_{\text{peak}}=0$. Notably, at these limit points, the domain created by eccentricity vanishes, resulting in the coincidence of the maximum and minimum values. Occasionally, the dynamic quantity value for the circular orbit may exceed the interpolation domain at $q=0.1$, situated on the boundary. In such instances, it is imperative to include it as the maximum or minimum value in the interpolation to uphold physical consistency, as the circular orbit's dynamic quantity value must be encompassed.

By comparing the interpolated domains of the two distinct scenarios, we observe that the domain shaped by $24.6M$ encompasses the domain delineated by $11.3M$, signifying that a larger initial distance engenders a broader boundary, aligning with our earlier analyses. Additionally, in FIG. \ref{FIG:7}, we incorporate scattered points representing circular orbit dynamics and their polynomial fits for a more intuitive comparison. Notably, the domain formed by eccentricity significantly surpasses the extent captured by the polynomial fit.

As previously noted, the data points for $24.6M$ are not densely distributed, leading to situations where the data points for $11.3M$ may fall outside the range of $24.6M$. To establish an upper limit for the domain influenced by the orbital eccentricity effect, a synthesis of the outcomes from both cases is imperative.

In the right panels (b), (d), (f), and (h) of FIG. \ref{FIG:7}, we showcase the domains derived from the combination of data points for $11.3M$ and $24.6M$. This combination involves determining the maximum or minimum values for the same mass ratio of $11.3M$ and $24.6M$—selecting the larger or smaller value accordingly. Additionally, when encountering data points that are notably sparse and exist solely for $11.3M$ but not for $24.6M$ (e.g., at $q=0.75$ and $D_{\text{ini}}=11.3M$), these data points are excluded. This approach ensures the derivation of a more rational and comprehensive domain, enhancing the overall integrity of the analysis.

In the right panels (b), (d), (f), and (h) of FIG. \ref{FIG:7}, the variations in dynamical quantities relative to the circular orbit induced by the orbital eccentricity effect are displayed, which is the upper limit of the current NR simulation, manifesting as domains enveloping the dynamical quantities of the circular orbital merger. In the future, there will be more comprehensive simulations of eccentricity BBH mergers to improve this upper limit. It is essential to quantitatively characterize the residual deviations of these domains compared to the circular orbit and elucidate how these residuals diverge from those of polynomial modeling for the dynamic quantities of the circular orbit.

\begin{figure*}[htbp!]
\centering
\includegraphics[width=16cm,height=8cm]{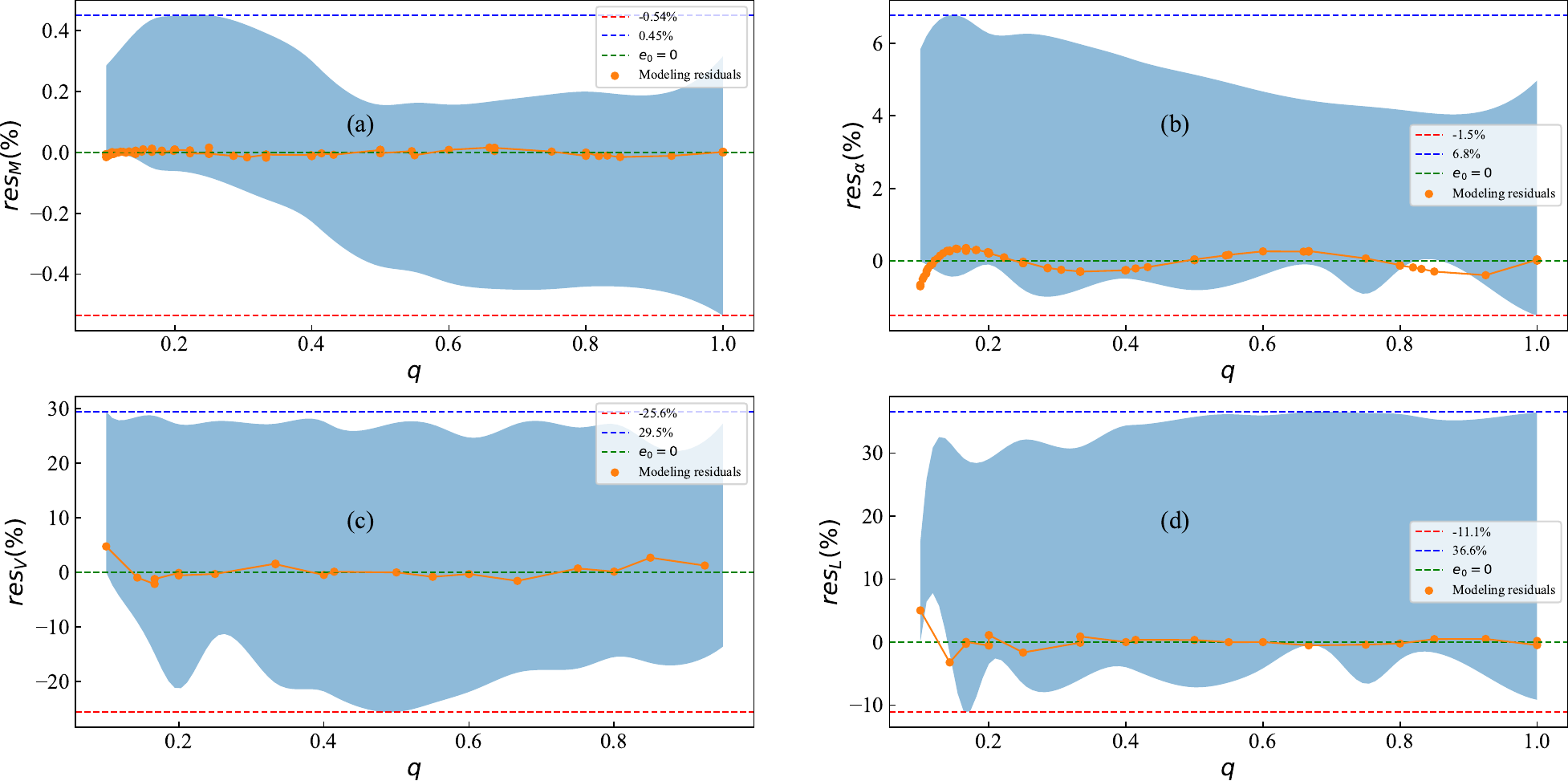}
\caption{\label{FIG:8}Residuals of the dynamic quantities $M_{\text{rem}}$, $\alpha_{\text{rem}}$, $V_{\text{rem}}$, and $L_{\text{peak}}$ resulting from the orbital eccentric effect relative to the dynamic quantities of the circular orbit (Blue shaded area). We include the residuals from the polynomial modeling of circular orbits for comparison (Modeling residuals).}
\end {figure*}

In FIG. \ref{FIG:8}, we employ Eq. (\ref{eq:20}) to compute the residuals of the dynamic quantities $M_{\text{rem}}$, $\alpha_{\text{rem}}$, $V_{\text{rem}}$, and $L_{\text{peak}}$ resulting from the eccentric effect relative to the dynamic quantities of the circular orbit. This computation is conducted based on the boundaries of these domains (i.e., the maximum and minimum values). When applying Eq. (\ref{eq:20}), we substitute $A_{\text{poly}}$ with the boundary values of the domains and indicate the label in the lower right corner of each residual to distinguish them in FIG. \ref{FIG:8}. To ensure consistency, we interpolate the circular orbit data during the calculations to align with the boundaries of the domains. We include the residuals from the polynomial modeling of circular orbits for comparison.

In FIG. \ref{FIG:8}, it is observed that the residuals stemming from the eccentric effect on the BBH merger are significantly larger than those derived from polynomial modeling of circular orbit. For instance, in panel (a), the biggest positive residuals from the $M_{\text{rem}}$ modeling reach up to 0.45\%, while the smallest negative residuals extend to -0.54\%. These positive and negative residuals arise due to the dynamic quantities in the domain surpassing or falling below the circular orbit data. In panels (b), (c), and (d) of FIG. \ref{FIG:8}, the biggest positive residuals from $\alpha_{\text{rem}}$, $V_{\text{rem}}$, and $L_{\text{peak}}$ modeling peak at 6.8\%, 29.5\%, and 36.6\%, respectively, with smallest negative residuals reaching -1.5\%, -25.6\%, and -11.1\%. The residuals introduced by these domains significantly surpass those of polynomial modeling. For $M_{\text{rem}}$, $\alpha_{\text{rem}}$, $V_{\text{rem}}$, and $L_{\text{peak}}$, the residuals stemming from the domains are approximately 25 times, 10 times, 6 times, and 7 times larger than the residuals of polynomial modeling for the circular orbit. This stark contrast is evident from the substantial disparity between the maximum and minimum values and the residuals of polynomial modeling in FIG. \ref{FIG:8}.

The correlations among eccentric dynamical quantities are notably intricate and challenging to generalize compared to their relationships with the mass ratio, often forming spiral or other complex structures. Nevertheless, as evidenced in FIG. \ref{FIG:6}, the dynamical quantities are all contingent on the mass ratio $q$, and concerning this relationship, they exhibit vertical oscillations solely. This characteristic enables us to delineate and extend the domains in FIG. \ref{FIG:7} to encompass any mass ratio.

\begin{figure*}[htbp!]
\centering
\includegraphics[width=16cm,height=12cm]{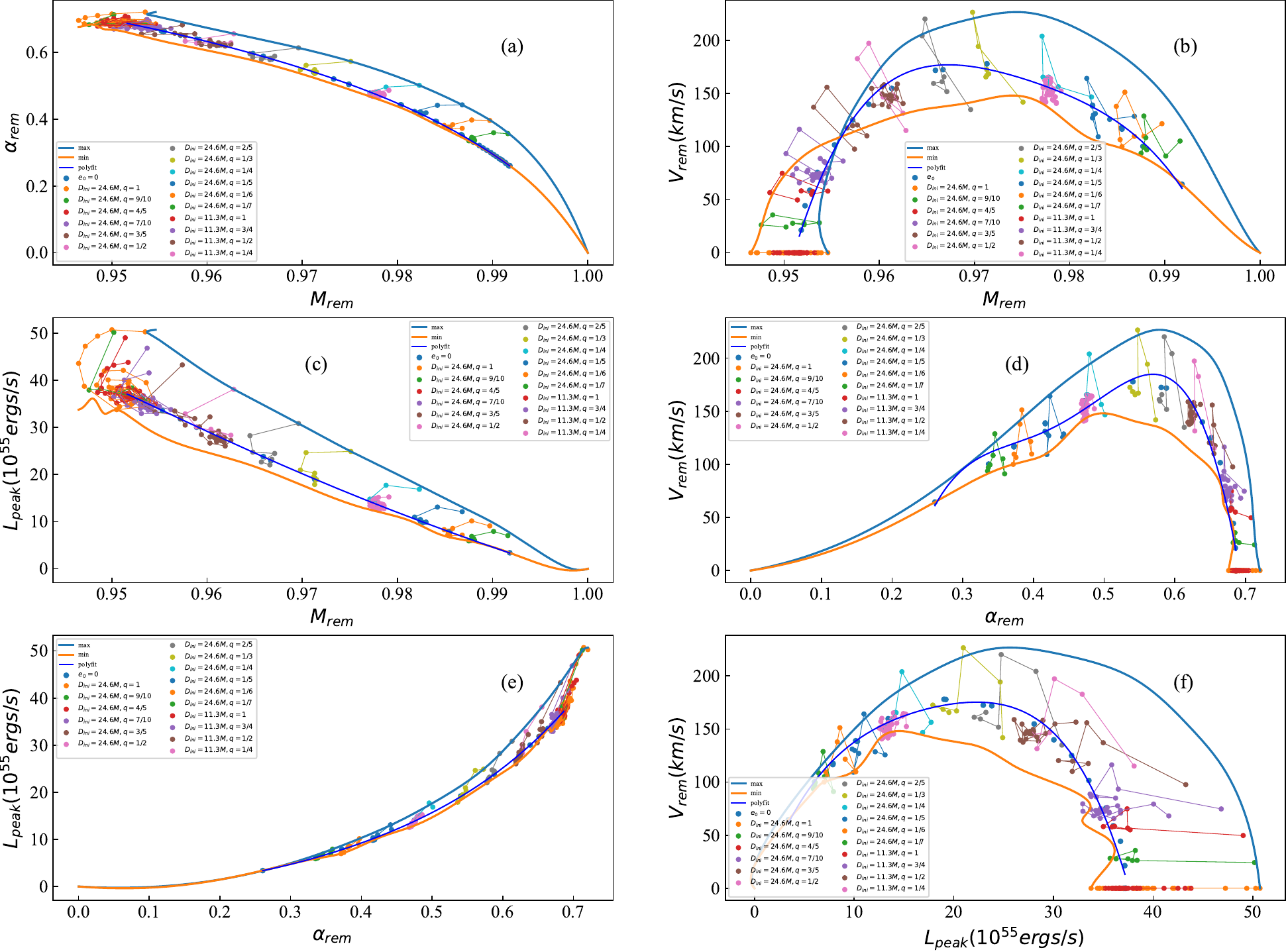}
\caption{\label{FIG:9}The maximum (max) and minimum (min) values from the relationships between the dynamic quantities and mass ratio constrain these correlations for the orbital eccentric BBH merger. We also include the polynomial modeling (polyfit) of circular orbits for comparison.}
\end {figure*}

In FIG. \ref{FIG:7}, we establish the maximum (upper boundary) and minimum (lower boundary) values of the dynamical quantities in panels (b), (d), (f), and (h). These extremities enable us to define boundaries for the correlations among the dynamical quantities. In FIG. \ref{FIG:9}, we impose constraints on the correlations among the dynamical quantities $M_{\text{rem}}$, $\alpha_{\text{rem}}$, $V_{\text{rem}}$, and $L_{\text{peak}}$ for the orbital BBH merger scenario utilizing the maximum and minimum values from panels (b), (d), (f), and (h) of FIG. \ref{FIG:7}.

The salience of this constraint lies in our ability to outline approximate domains for the correlations among the dynamical quantities engendered by the eccentricity effect, despite lacking detailed insights into the specific structures of these dynamic quantities. FIG. \ref{FIG:9} adeptly captures how the maximum and minimum values from the relationships between the dynamic quantities and mass ratio effectively constrain these correlations for the orbital eccentric BBH merger, enabling an evaluation of the orbital eccentricity effect and the residual magnitude compared to the polynomial fit based on the depicted curves. Notably, in panel (b) of FIG. \ref{FIG:9}, the intersection of the maximum and minimum values at the midpoint signifies a more intricate correlation in this particular case, but the boundaries also impose certain constraints on this correlation.

\subsection{Non-orbital merger}\label{sec:III:B}
\begin{figure*}[htbp!]
\centering
\includegraphics[width=16cm,height=8cm]{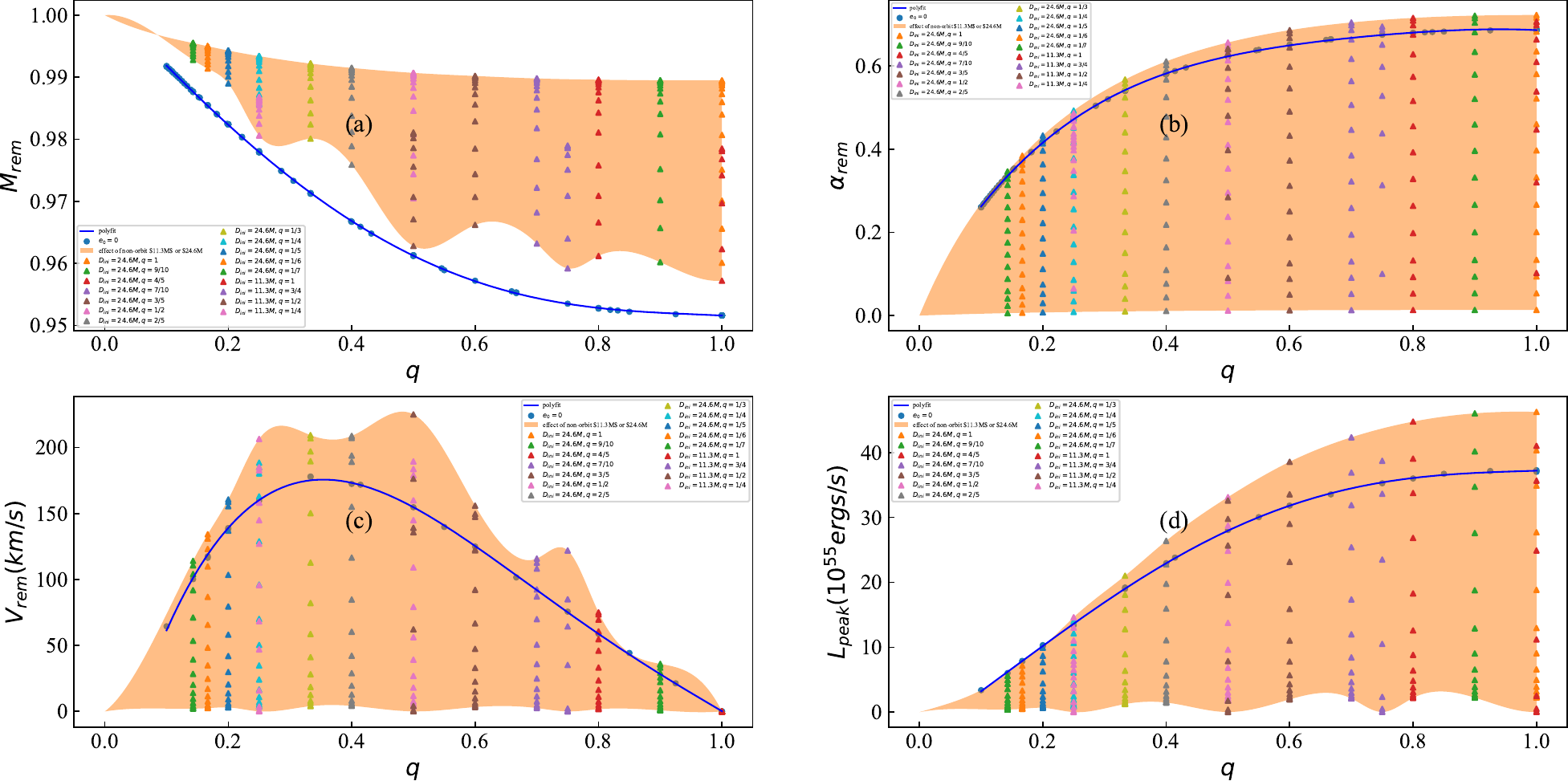}
\caption{\label{FIG:10}Domains of the dynamical quantities influenced by non-orbital eccentricity effects relative to circular orbits. We include the polynomial modeling (polyfit) of circular orbits for comparison.}
\end {figure*}

The opposite of orbital mergers is non-orbital mergers, wherein the BBH fails to complete a full orbital cycle. Consequently, the dynamical quantities $M_{\text{rem}}$, $\alpha_{\text{rem}}$, $V_{\text{rem}}$, and $L_{\text{peak}}$ gradually tend towards a stable value or zero, as depicted in FIG. \ref{FIG:6}. Similar to the boundaries (maximum and minimum) derived from the relationships between the dynamical quantities and the mass ratio in the panels (b), (d), (f), and (h) on the right side of FIG. \ref{FIG:7}, FIG. \ref{FIG:10} depicts the delineated domains of dynamical quantities influenced by eccentricity effects in comparison to circular orbits. These domains are established through the interpolation of maximum and minimum values of dynamical quantities in non-orbital mergers, representing an upper limit derived from current NR data. To streamline the presentation, we omit the comparison between the two initial distances in this context, akin to the scenarios depicted on the left side of FIG. \ref{FIG:7}.

The domains delineated in FIG. \ref{FIG:10} appear notably larger and broader compared to those associated with orbital BBH mergers in FIG. \ref{FIG:7}. Notably, the domain in panel (a) of FIG. \ref{FIG:10} does not encompass the dynamical quantities of the circular orbit, while the domains in panels (b), (c), and (d) do include these quantities. The residuals arising from these domains significantly surpass the residuals derived from polynomial modeling of the circular orbit. These domains extend to the boundaries of the parameter space for $M_{\text{rem}}$, $\alpha_{\text{rem}}$, $V_{\text{rem}}$, and $L_{\text{peak}}$—a testament to the prevalence of non-orbital mergers.

\begin{figure*}[htbp!]
\centering
\includegraphics[width=16cm,height=12cm]{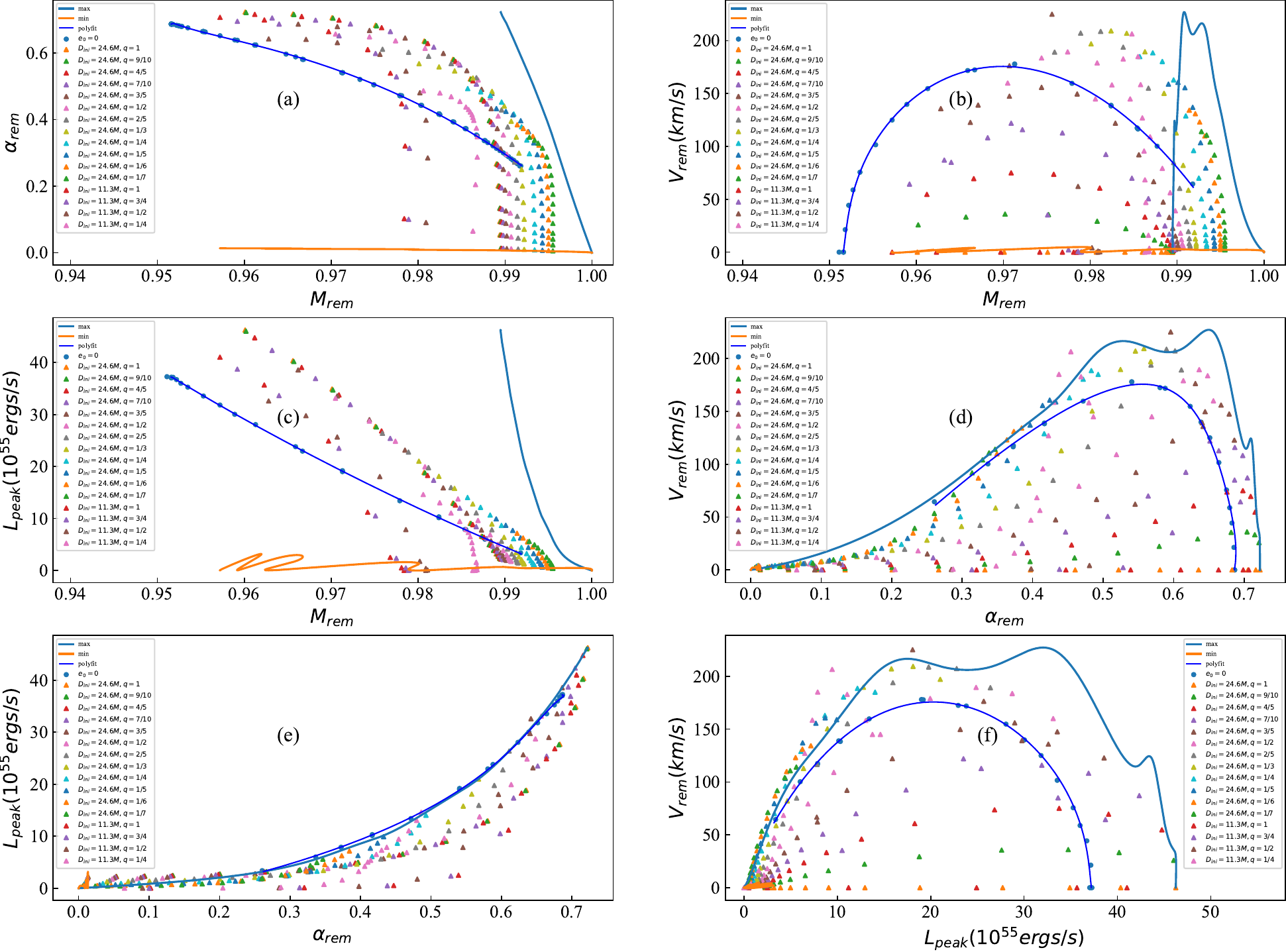}
\caption{\label{FIG:11}The maximum (max) and minimum (min) values from the relationships between the dynamic quantities and mass ratio constrain these correlations for the non-orbital eccentric BBH merger. We include the polynomial modeling (polyfit) of circular orbits for comparison.}
\end {figure*}

In FIG. \ref{FIG:11}, akin to FIG. \ref{FIG:9}, The maximum (max) and minimum (min) values from the relationships between the dynamic quantities and mass ratio are presented as constraints on the correlations among the dynamical quantities of non-orbital mergers. Unlike FIG. \ref{FIG:9}, the constraints in FIG. \ref{FIG:11} are not rigid, occasionally deviating significantly from the scatter points. Furthermore, some minimum boundaries collapse into narrow line segments, offering partial insights into the actual lower boundary of the dynamical quantities. The intersection of the maximum and minimum of penal (b) is stronger. We also include the polynomial modeling of circular orbits for comparison.

\section{Extension to general BBH configuration}\label{sec:IV}
For nonspinning BBHs, both the orbital and non-orbital effects of eccentricity give rise to domains where the relationships between the dynamical quantities $M_{\text{rem}}$, $\alpha_{\text{rem}}$, $V_{\text{rem}}$, and $L_{\text{peak}}$ and the mass ratio $q$. And the relationships form boundaries that constrain their correlations concerning their corresponding circular orbits. This unique role of eccentricity is paramount.

Does this effect persist in the context of spin-aligned and spin-precessing BBH mergers? The answer is affirmative. However, a comprehensive validation necessitates more extensive BBH simulations focusing on the eccentric orbits of these two configurations.

This effect emanates from eccentricity inducing oscillations in the amplitude of gravitational waves, impacting all phases of BBH motion: inspiral, merger, and ringdown. Modifying the initial mean anomaly $l_0$ enables the alteration of oscillation patterns, ultimately leading to the formation of an envelope that encapsulates all oscillations, as deduced in Sec. \ref{sec:II:C}. The oscillation of gravitational wave amplitude in eccentric orbital BBHs is pivotal in generating dynamic oscillations and forming domains.

The waveform amplitude of spin-aligned and spin-precessing BBHs also universally oscillates in the presence of eccentricity. The waveform amplitude of spin-aligned BBHs mirrors that of nonspinning BBHs, showing oscillations caused solely by eccentricity. In contrast, the waveform amplitude of spin-precessing BBHs oscillates due to both precession and eccentricity. Notably, precession-induced oscillations are distinct from those induced by eccentricity and operate independently of them. Previous analyses in Ref. \cite{Wang:2023vka} elucidate that the dynamical quantities of eccentric orbit BBHs with spin alignment and spin precession exhibit a near-horizontal relationship with small initial eccentricity. At higher eccentricities, these quantities reach peaks or valleys primarily due to the influence of eccentricity, a consistency observed in the nonspinning case. Hence, BBH mergers characterized by eccentric orbits with spin alignment and spin precession also manifest this phenomenon.

\section{Conclusion and Outlook}\label{sec:V}
In this paper, we firstly revisit and analyze the oscillation phenomenon of radiative quantities—energy, angular momentum, and linear momentum—associated with initial eccentricities, as discussed in our earlier work \cite{Wang:2024afj}. We find that by continuously varying the mean anomaly $l_0$ within the parameter space $[0,2\pi]$, we can form an envelope that encapsulates the original oscillations of these radiative quantities. This observation indicates that the oscillations arise from the specific initial condition $l_0$, while the influence of the actual eccentricity contributes to the formation of the envelope.

Next, we model the peak luminosity $L_{\text{peak}}$, mass $M_{\text{rem}}$, spin $\alpha_{\text{rem}}$, and recoil velocity $V_{\text{rem}}$ in circular orbits, exploring their relationships with the mass ratio using polynomial modeling. Additionally, we examine their mutual correlations through the same modeling approach and find that polynomials effectively capture these relationships.

We then synthesize and analyze the dynamical quantities for both circular and eccentric orbits. We conclude that these quantities should exhibit continuous variation within specific ranges dictated by mass ratios, influenced by the continuous changes in initial eccentricity and the associated envelope. This understanding can be extended to other mass ratios. Consequently, in the context of orbital and non-orbital mergers, we interpolate the maximum and minimum values of the dynamical quantities to delineate the domain of dynamical quantities for eccentric BBH mergers. This domain is significantly broader than the polynomial modeling of circular orbits, providing constraints on the correlations among dynamical quantities of eccentric mergers. As the underlying mechanism for this effect is linked to oscillations in gravitational wave amplitude due to eccentricity, our findings can also be applied to scenarios involving spin alignment and precession.

The domains of dynamical quantities shaped by eccentricity in relation to circular orbits are derived from all current eccentric BBH simulations conducted by RIT, and they are not exhaustive. These domains represent the upper limits of our current understanding, and we anticipate that their extents will expand with increased initial separations in eccentric NR simulations. As these domains develop, our comprehension of eccentric BBH mergers will improve, with future simulations at larger initial distances likely to refine the boundaries of these domains. Moreover, further simulations incorporating spin alignment and spin precession configurations will enhance our understanding of eccentric orbital BBH mergers.

\begin{acknowledgments}
The authors are very grateful to the RIT and SXS collaboration for the numerical simulation of BBH mergers, and thanks to Yan Fang Huang, Yu Liu and Xiaolin Liu for their helpful discussions. The computation is partially completed in the HPC Platform of Huazhong University of Science and Technology. This work is supported by the National Key R\&D Program of China (2021YFA0718504).
\end{acknowledgments}

\bibliographystyle{apsrev4-2}

\bibliography{ref}

\clearpage

\end{document}